\documentclass[aip,jcp,reprint,superscriptaddress,floatfix]{revtex4-2}

\usepackage{amsmath}
\usepackage{amssymb}
\usepackage{graphicx} 
\usepackage{booktabs}
\usepackage{comment}
\usepackage{braket}
\usepackage[english]{babel}
\usepackage{hyperref}

\usepackage[version=4]{mhchem}

\usepackage[usenames,dvipsnames]{xcolor}
\usepackage{soul}

\begin{document}
\include{defs}

\title{Synthetic pre-training of graph-network models for predicting solid-state NMR parameters}

\author{Chiheb Ben Mahmoud}
\email{chiheb.benmahmoud@materials.ox.ac.uk}
\affiliation{Department of Materials, University of Oxford, Oxford OX1 3PH, United Kingdom}
\author{Carlos Bornes}
\affiliation{Department of Physical and Macromolecular Chemistry, Charles University, Hlavova 8, Praha 2, Prague 12800, Czech Republic}
\author{Christopher J. Heard}
\affiliation{Department of Physical and Macromolecular Chemistry, Charles University, Hlavova 8, Praha 2, Prague 12800, Czech Republic}
\author{Lukáš Grajciar}
\affiliation{Department of Physical and Macromolecular Chemistry, Charles University, Hlavova 8, Praha 2, Prague 12800, Czech Republic}
\author{Jonathan R. Yates}
\affiliation{Department of Materials, University of Oxford, Oxford OX1 3PH, United Kingdom}
\author{Volker L. Deringer}
\affiliation{Inorganic Chemistry Laboratory, Department of Chemistry, University of Oxford, Oxford OX1 3QR, United Kingdom}

\date{\today}

\begin{abstract}
    Nuclear magnetic resonance (NMR) is a powerful probe of atomic structure, but accurate quantum-mechanical predictions of tensorial NMR parameters are computationally demanding. This creates a bottleneck both for direct quantum-mechanical studies and for collecting high-quality training data for machine-learning (ML) models.
    Here, we introduce a synthetic pre-training and fine-tuning protocol for graph-based ML models of solid-state NMR parameters. We first pre-train models on synthetic tensorial data, as obtained using an existing ML model, and subsequently fine-tune those models on new ground-truth data. We observe a pronounced improvement in data efficiency when pre-training and fine-tuning span the same compositional and configurational space, and we carry out initial experiments regarding chemical transferability.
    Our work outlines a route toward future data-efficient training workflows for tensorial ML models for solid-state NMR, combining inexpensive synthetic supervision with targeted first-principles refinement.

\end{abstract}

\maketitle

\section{Introduction}

Nuclear magnetic resonance (NMR) spectroscopy is a powerful probe of local atomic structure in molecules and materials~\cite{kuhn_nuclear_2024,das_exploring_2025}. In solid-state NMR, the magnetic shielding (MS) and electric field gradient (EFG) tensors are central quantities governing experimental observables such as powder line shapes and quadrupolar broadening, which cannot be reproduced from isotropic shifts alone. Accurate prediction of these tensorial parameters from quantum-mechanical (QM) computations is therefore essential for interpreting experimental spectra and validating structural models, motivating the development of methodology for efficient tensorial NMR predictions.

Machine learning (ML) approaches provide an efficient route to predicting NMR parameters in molecular and materials systems~\cite{rupp_machine_2015,cuny_initio_2016,paruzzo_chemical_2018,chaker_nmr_2019,gerrard_impression_2020,kwon_neural_2020,han_scalable_2022,venetos_machine_2023,bankestad_carbohydrate_2024,charpentier_firstprinciples_2025,harper_performance_2025}. However, these ML models face two major challenges. The first is representing the rotational symmetry of NMR tensors, since their components transform nontrivially under rotations, in contrast to scalar learning targets such as energies or isotropic shifts. Equivariant ML architectures, including approaches based on symmetry-adapted descriptors~\cite{grisafi_symmetryadapted_2018} or geometric graph neural networks (GNNs)~\cite{batzner_e3equivariant_2022,batatia_mace_2022,duval_hitchhikers_2024}, address this symmetry constraint by enforcing the correct transformation properties of tensorial outputs and by providing a physically motivated inductive bias. Both classes of methods have been successfully applied to the prediction of tensorial NMR parameters~\cite{venetos_machine_2023,charpentier_firstprinciples_2025,harper_performance_2025,benmahmoud_graphneuralnetwork_2025,kellner_deep_2025}. The second major challenge is data availability: these models rely on reference QM data obtained using methods such as the gauge-independent atomic orbital (GIAO)~\cite{ditchfield_selfconsistent_1974,wolinski_efficient_1990} for molecules or the gauge-including projector augmented wave (GIPAW)~\cite{pickard_allelectron_2001,yates_calculation_2007,charpentier_paw_2011,bonhomme_firstprinciples_2012} for periodic systems. The cost of well-converged tensorial NMR parameters, together with the uneven distribution of local structural motifs, limits the availability of training data, particularly for chemically and structurally heterogeneous materials.

Much current research in atomistic ML is focused on the development of large-scale ``foundational'' models, particularly for ML interatomic potentials (MLIPs)~\cite{deng_chgnet_2023,neumann_orb_2024,batatia_foundation_2025,mazitov_petmad_2025}, where pre-training on diverse datasets yields transferable models for energies and forces. Extending this paradigm to tensorial NMR parameters in materials is more challenging~\cite{bornes_accurate_2026}, both because of the computational cost and because NMR parameters are not available for certain classes of systems or configurations (say, metallic snapshots). Transferable NMR models for molecular solids provide an important step in this direction~\cite{kellner_deep_2025}.

A promising approach to address data scarcity in ML is the use of \emph{synthetic data}. In the context of the present work, we take the term to mean training labels generated using approximate or auxiliary models, including lower-accuracy methods or pre-existing ML~\cite{gardner_synthetic_2023} and empirical models~\cite{NEURIPS2022_5ef1df23}. Synthetic labels have been leveraged to increase the data efficiency of MLIPs~\cite{gardner_synthetic_2024}, where models pre-trained on synthetic data and subsequently fine-tuned on a limited number of QM labels achieve accuracies comparable to those trained on substantially larger QM datasets. It was also shown that pre-training increases robustness in molecular-dynamics (MD) simulations, both for DFT-level~\cite{zaverkin_transfer_2023} and synthetic~\cite{gardner_synthetic_2024} pre-training. Related approaches use large and diverse datasets and multi-task pre-training to improve model accuracy, suggesting a broader trend toward richer training regimes beyond direct (pre-) training on QM data~\cite{zhang_dpa2_2024,benmahmoud_data_2024}. 

Here, we propose a pre-training and fine-tuning approach for graph-based ML models of tensorial NMR parameters. We show how synthetic labels can be integrated into standard training protocols in this area, making it possible to expand structural and chemical coverage and reduce reliance on large QM training sets. 
We examine this approach in two settings. First, we assess it for an ``in-domain'' scenario: pre-training and fine-tuning a model on first-principles MS tensors in amorphous silica (a-\ce{SiO2}). Second, we evaluate the transferability of the pre-trained model to a more chemically diverse system, viz.\ Si--O--Al--H-based zeolites.
Our findings provide proof-of-concept that synthetic pre-training can improve training efficiency for NMR parameters, thereby reducing the computational cost associated with generating high-quality training data. 

\section{Methods}

\subsection{Graph-network models for NMR parameters}

We build upon the MACE~\cite{batatia_mace_2022} framework, as a representative graph-based architecture, to model local atom-centered tensorial NMR parameters, in particular the MS tensor $\boldsymbol{\sigma}$. MACE represents a chemical structure as a graph: nodes carry atomic species information and edges encode local geometric relations within a finite radial cutoff. Atomic environments are expanded in radial and angular basis functions, and the atomic features are iteratively refined through message passing.

MACE is implemented as an E(3)-equivariant model~\cite{batatia_design_2025}, so that internal features transform according to irreducible representations of the rotation group. In the standard MACE implementation, the available equivariant channels follow an alternating parity pattern with angular momentum, such that the parity is given by $(-1)^{\ell}$, where $\ell$ is the angular momentum quantum number. This means that channels with even $\ell$ have even parity and those with odd $\ell$ have odd parity. 

The spherical representation of $\boldsymbol{\sigma}$ is written as
\begin{equation}
    \boldsymbol{\sigma} = \sigma^{(0)} \oplus \sigma^{(1)} \oplus \sigma^{(2)},
\end{equation}
where $\sigma^{(0)}$ represents the isotropic contribution, taken here to be the average of the trace of the MS tensor, $\sigma^{(1)}$ represents the antisymmetric part, and $\sigma^{(2)}$ represents the traceless symmetric part. 
The Haeberlen asymmetry $\sigma_{\eta}$ and anisotropy $\sigma_{\zeta}$ parameters~\cite{haeberlen_advances_1976} are computed from the reconstructed Cartesian  MS tensor and depend on $\sigma^{(2)}$. These quantities characterize the NMR line shape for spin-1/2 nuclei. They are also commonly reported experimental observables and used as inputs for NMR simulation codes. 
$\sigma^{(0)}$ and $\sigma^{(2)}$ can be learned directly by the current MACE architecture, while $\sigma^{(1)}$ is challenging because of the parity mismatch with corresponding MACE features. 

To solve this practical problem, we construct the target rank-2 spherical tensor by considering a linear combination of couplings of  equivariant outputs of arbitrary rank, as introduced in Ref.~\citenum{benmahmoud_graphneuralnetwork_2025}. In particular, a set of equivariant outputs \{$T_i^{(\ell)}$, $T_j^{(\ell)}$\} transforming according to the $\ell$-th irreducible representation and sharing the same parity can be combined via the standard angular-momentum coupling coefficients:
\begin{equation}\label{eq:decomp}
    \sum_{i=1}^N \sum_{j=1}^N w_{i,j}~T_i^{(\ell)} \otimes T_j^{(\ell)} = \bigoplus_{L = 0}^{2\ell} T_i^{(L)},
\end{equation}
where $N^2$ is the number of tensors involved in the coupling.  
This formulation provides the model with sufficient expressivity to  represent all components while preserving the rotational covariance. This construction yields a general spherical tensor representation of the target that is fully symmetry-consistent, irrespective of the single-channel parity constraints natively supported by the existing implementations in \texttt{mace-torch}~\cite{batatia_mace_2026} and \texttt{graph-pes}~\cite{gardner_graph_2026}.

We implement the MACE-based model architecture discussed in this work in the \texttt{graph-pes} package, and refer to it as \texttt{TensorMACE},  to emphasize its extension to atom-centered tensorial predictions, such as the MS. We use this architecture with the linear combination of tensor products from Eq.~\ref{eq:decomp}. In this work, the \texttt{TensorMACE} model targets the MS tensor $\boldsymbol{\sigma}$, represented through its spherical tensor decomposition into components $\sigma^{(0)}$, $\sigma^{(1)}$, and $\sigma^{(2)}$, which transform according to irreducible representations of the rotation group. After prediction, we reconstruct the Cartesian MS tensor and compute the asymmetry $\sigma_{\eta}$, and the anisotropy $\sigma_{\zeta}$ parameters.
 
\subsection{Datasets}

We carry out numerical experiments using three datasets. The first consists of 5000 a-\ce{SiO2} structures generated under varying conditions following the protocol detailed in Refs.~\citenum{erhard_modelling_2024,benmahmoud_graphneuralnetwork_2025}. Each configuration contains 144 atoms. The MS tensors of these structures are generated using a model based on the NequIP architecture~\cite{batzner_e3equivariant_2022,benmahmoud_assessing_2025} trained on 211 a-\ce{SiO2} snapshots generated with the same protocol. The main hyperparameters of this synthetic-label generator are a radial cutoff of 5.53 \AA{}, 6 message-passing layers and 128 hidden features per equivariant rank (up to rank $L=2$) per layer. 
This model yields moderate accuracy: the RMSE calculated for the isotropic MS term $\sigma^{(0)}$ of silicon is approximately $1$~ppm and of oxygen is $\approx 2.1$~ppm in the a-\ce{SiO2} test. 

The second dataset, taken from Refs.~\citenum{benmahmoud_graphneuralnetwork_2025, benmahmoud_graphneuralnetwork_2025a}, consists of 450 amorphous silica (a-\ce{SiO2}) structures generated under the same conditions as the first dataset. 
Each configuration contains 144 atoms. 

The third dataset is assembled as a subset of the zeolitic structures reported in Refs.~\citenum{erlebach_reactive_2024,lei_machine_2025}. It comprises chemically more diverse Si--O--Al--H environments, including elements that are absent from the a-\ce{SiO2} dataset. The configurations are drawn from MD trajectories of H-exchanged aluminosilicate zeolites spanning ten topologies, three known (CHA, SOD, MVY)~\cite{mccusker_nomenclature_2001,baerlocher_database_} and seven hypothetical, with Si/Al ratios from 1 to 32 and water loading from dry to around 1.1 g cm$^{-3}$. We filter the configurations containing the elements Si, O, Al, and H and choose an initial pool of 1000 structures chosen by Farthest Point Sampling~\cite{imbalzano_automatic_2018}. From this subset, we select 450 configurations randomly. The resulting configurations range from 6 to 192 atoms (2 to 36 \ce{SiO2} formula units). 
In the previous two datasets, 50 structures serve as a validation set and 150 structures as a test set. 

We compute the MS tensors for the second and third datasets using the GIPAW formalism as implemented in CASTEP~\cite{profeta_accurate_2003,clark_first_2005,yates_calculation_2007}  together with on-the-fly-generated pseudopotentials. For the a-\ce{SiO2} dataset, we use the Perdew--Burke--Ernzerhof (PBE)~\cite{perdew_generalized_1996} exchange--correlation functional, a fine $k$-point spacing of $0.03$~\AA$^{-1}$, and a plane-wave energy cutoff of $900$~eV.
For the Si--O--Al--H dataset, we use a solid-optimized version of the PBE functional, denoted PBEsol ~\cite{perdew_restoring_2008}, which generally yields improved accuracy for structural parameters of solids compared to PBE~\cite{csonka_assessing_2009,tran_rungs_2016}. We also use a fine $k$-point spacing of $0.05$~\AA$^{-1}$, and a plane-wave energy cutoff of $900$~eV. We choose these parameters to ensure convergence of the MS tensors for all considered atoms. 

\begin{figure*}
    \centering
    \includegraphics[scale=1]{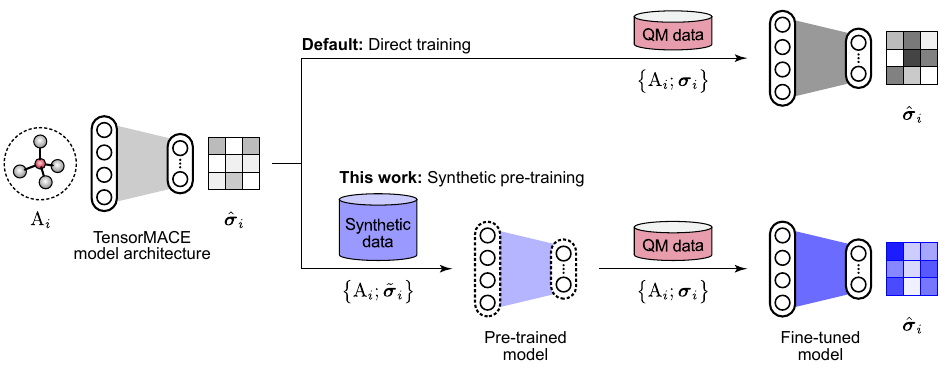}
    \caption{Synthetic pre-training for graph-based models of tensorial NMR parameters. 
    The \texttt{TensorMACE} architecture ({\em left}) maps atomic environments, A$_{i}$, onto predictions of tensorial quantities for that atom, $\hat{\boldsymbol{\sigma}}_i$.
    Instead of directly training models on QM NMR data $\boldsymbol{\sigma}_i$ ({\em top right}), we pre-train models on a synthetic dataset containing approximate tensorial targets $\tilde{\boldsymbol{\sigma}}_i$ and subsequently fine-tune on $\boldsymbol{\sigma}_i$ ({\em bottom right}). 
    Adapted in parts from Refs.~\citenum{benmahmoud_graphneuralnetwork_2025} and \citenum{gardner_synthetic_2024}, which are both published under the CC BY 4.0 license (\href{https://creativecommons.org/licenses/by/4.0/}{https://creativecommons.org/licenses/by/4.0/})
    }
    \label{fig:workflow}
\end{figure*}

\subsection{Model fitting}\label{sec:fitting}

We consider two different \texttt{TensorMACE} architectures that differ by their size: we call them ``large'' (L) and ``small'' (S), respectively. Both share the same cutoff (6.0~\AA), number of message-passing layers (3), channel width (160), and number of tensor products (4096). They differ only in the maximum feature rank (3 for the large model and 1 for the small model) and therefore the output rank before the coupling, thereby isolating the effect of tensorial expressivity on model performance. 

We train \texttt{TensorMACE} models by minimizing a mean square error loss on the spherical components, including a learnable per-element offset. In practice, we express the predicted MS tensor $\hat{\boldsymbol{\sigma}}_i$ as:
\[
    \hat{\boldsymbol{\sigma}}_i = \hat{\boldsymbol{\sigma}}_i^{\mathrm{\texttt{TensorMACE}}} + \mathbf{b}_{\alpha(i)},
\]
where $\mathbf{b}_{\alpha(i)}$ contains the learnable offsets associated with the chemical species $\alpha(i)$. $\mathbf{b}_{\alpha(i)}$ represents the mean of the spherical contributions over the training set per chemical species. 
We use batch sizes between 4 and 10, depending on the size of the training dataset. We use the AdamW~\cite{loshchilov2018decoupled} optimizer with an initial learning rate of 0.003. We reduce the learning rate adaptively using a scheduler with a patience of 10 epochs and apply early stopping if the validation error does not improve for 100 consecutive epochs. The models typically converge within approximately 200 epochs. We train four instances of each model using a different random seed, and report the average of their prediction error.

To compare the trained models across the different NMR-related properties, we use the relative root mean square error (\%RMSE), defined as the root-mean-square error (RMSE) normalized by the corresponding standard deviation and expressed as a percentage:
\begin{equation*}
    \%\mathrm{RMSE} = 100\times \dfrac{\sqrt{\sum_i (y^{\mathrm{ML}}_i - y^{\mathrm{QM}}_i)^2}}{\sqrt{\sum_i (y^{\mathrm{QM}}_i - \bar{y}^{\mathrm{QM}})^2}},
\end{equation*}
where $y^{\mathrm{ML}}_i$ is the ML model's prediction of $\sigma^{(0)}$, $\sigma^{(1)}$, and $\sigma^{(2)}$, or the Haeberlen parameters, and $\bar{y}^{\mathrm{QM}}$ is the average of the $y^{\mathrm{QM}}$ values across the dataset.

\section{Results and discussion}

\subsection{Synthetic pre-training workflow}
\begin{figure*}
    \centering
    \includegraphics[width=1.\linewidth]{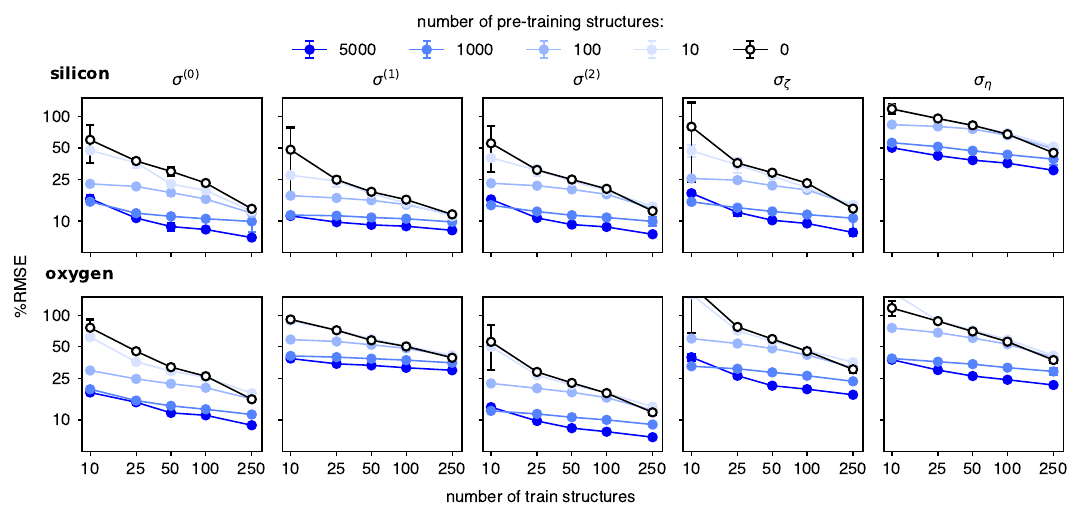}
    \caption{Evolution of errors in synthetically-pre-trained-and-fine-tuned and directly-trained \texttt{TensorMACE} models for a-\ce{SiO2}. We show the prediction errors for isotropic and anisotropic components of the MS tensor for silicon and oxygen as a function of the QM NMR training structures.
    Shades of blue indicate the number of synthetic pre-training structures. Empty circles correspond to models trained directly.
    Error bars indicate the standard deviation of the prediction errors across four models trained with different random seeds. When not visible, the standard deviation is smaller than the marker size.} 
    \label{fig:sio2}
\end{figure*}

Figure \ref{fig:workflow} summarizes the synthetic pre-training strategy used in this work. We train \texttt{TensorMACE} models on a large dataset of atomic environments, $A_i$, labeled with synthetic tensorial targets $\tilde{\boldsymbol{\sigma}}_i$ (we use the tilde, $\sim$, symbol to indicate synthetic data, and a ``hat'' to indicate the final ML model predictions). Then, we fine-tune the model on high-quality QM tensors, $\boldsymbol{\sigma}_i$. For comparison, Fig.~\ref{fig:workflow} also shows the conventional \emph{direct} approach, where the model is randomly initialized and trained exclusively on QM data. We note that this workflow is agnostic to the GNN architecture, and is compatible with any neural-network-based models in principle. 

Because the GNN architecture and the loss function are identical in the two protocols, this setup allows us to study the effect of synthetic pre-training in isolation. Any improvement observed after fine-tuning, relative to the direct approach, can therefore be attributed to the synthetic data biasing the tensorial ML model toward learning correlations between local geometry and NMR parameters, and thereby initializing the model in a physically meaningful region of the parameter space.

To distinguish between training protocols, we use the following notation for trained models:  \texttt{TensorMACE}-$P$-$F$(size), where $P$ and $F$ are the number of pre-training and fine-tuning structures, respectively. For example, \texttt{TensorMACE}-5000-10(L) denotes a large model pre-trained on 5000 synthetic structures and fine-tuned on 10 QM-labeled ones; directly trained models correspond to $P=0$.

\subsection{Synthetic pre-training for a-SiO$_2$}\label{sec:pre-train-sio2}

\begin{figure*}
    \centering
    \includegraphics[width=1.\linewidth]{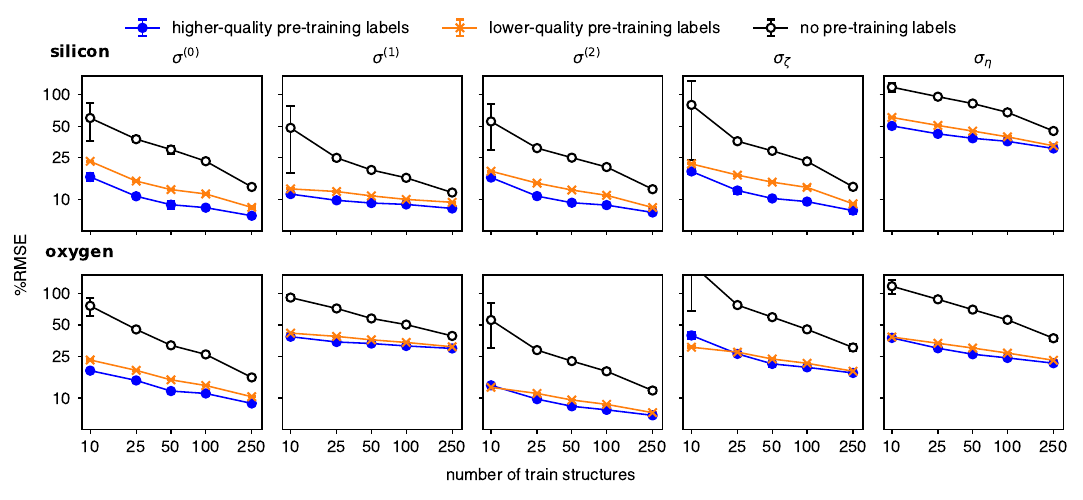}
    \caption{Evolution of errors in \texttt{TensorMACE}-5000-$F$(L) pre-trained using synthetic labels of different quality in a-\ce{SiO2}. We show the prediction errors for isotropic and anisotropic components of the MS tensor for silicon and oxygen as a function of the QM NMR training structures.
    Blue circles and oranges crosses correspond to models pre-trained with higher- and lower-quality synthetic labels, respectively; empty black circles denote directly trained \texttt{TensorMACE}-0-F(L) models. Error bars indicate the standard deviation of the prediction errors across four models trained with different random seeds. When not visible, the standard deviation is smaller than the marker size.}
    \label{fig:noisy}
\end{figure*}

We first assess the impact of synthetic pre-training in a setting where both the synthetic pre-training/fine-tuning and direct approaches are applied to an a-\ce{SiO2} dataset. Figure~\ref{fig:sio2} shows the evolution of the prediction error as a function of the number of QM-labeled structures in the training set, for the spherical components of the MS tensor and associated anisotropy parameters for silicon and oxygen, respectively. In these plots, the dark blue circles correspond to the pre-trained and then fine-tuned model, while the empty circles denote the directly trained model. We consider training set sizes ranging between 10 and 250 structures, where smaller sets are subsets of the larger ones. All models are tested on the same independent test set containing 150 configurations.

For all quantities considered across both chemical species, models initialized via synthetic pre-training achieve systematically lower errors than the corresponding directly-trained models when both are trained (or fine-tuned, in the case of the synthetically pre-trained models) on the same QM dataset, indicating a consistent improvement in data efficiency. The improvement is most pronounced when the training set contains fewer than 100 structures. As the size of the training dataset increases, the performance of the two models converges, indicating that synthetic pre-training primarily accelerates convergence toward a reasonable mapping rather than improving the ``asymptotic'' accuracy. Notably, for the particular case of $\sigma_\eta$, \texttt{TensorMACE}-5000-10(L) attains an accuracy comparable to a directly trained model using 250 structures. The accuracy of the anisotropy parameters is directly linked to the quality of the learned $\sigma^{(2)}$, which encodes orientation-dependent features of the tensor. In absolute terms, the fine-tuned model \texttt{TensorMACE}-5000-250(L) achieves average RMSE values of 0.6~ppm (silicon) and 1.58~ppm (oxygen) for $\sigma^{(0)}$, compared to 1.1~ppm and 2.8~ppm for the direct model \texttt{TensorMACE}-0-250(L). For $\sigma^{(2)}$, the corresponding errors are 1.3~ppm (silicon) and 1.9~ppm (oxygen) for the fine-tuned model versus 3.2~ppm and 3.3~ppm for the direct model.

\begin{figure*}
    \centering
    \includegraphics[width=1.\linewidth]{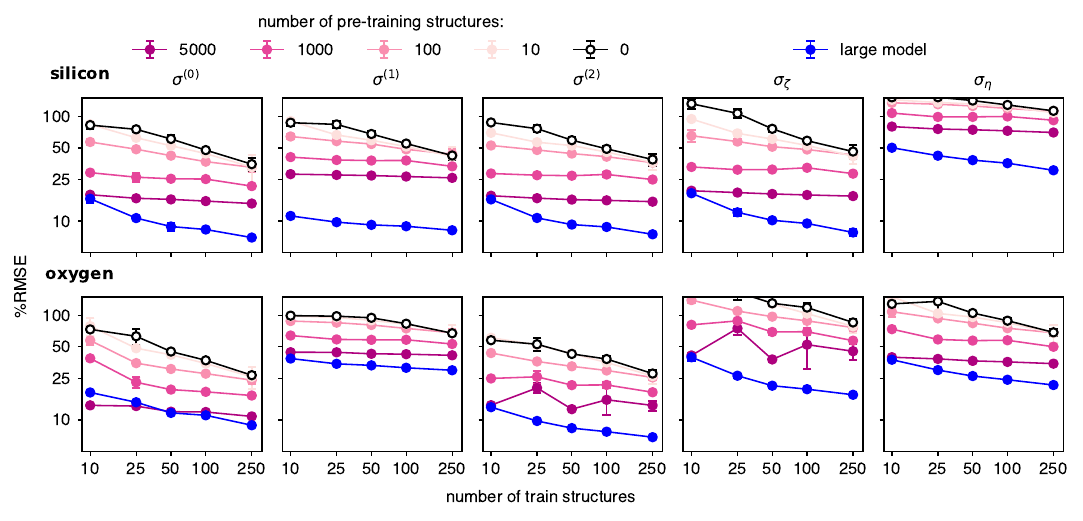}
    \caption{Evolution of errors in synthetically-pre-trained-and-fine-tuned and directly-trained small \texttt{TensorMACE} models in a-\ce{SiO2}. We show the prediction errors for isotropic and anisotropic components of the MS tensor for silicon and oxygen as a function of the QM NMR training structures.
    Shades of red indicate the number of synthetic pre-training structures. Empty circles correspond to models trained directly. Blue symbols refer to the best-performing synthetically-pre-trained-and-fine-tuned large \texttt{TensorMACE} model. Error bars indicate the standard deviation of the prediction errors across four models trained with different random seeds. When not visible, the standard deviation is smaller than the marker size.}
    \label{fig:sio2_small}
\end{figure*}

To investigate the role of synthetic data, we perform ablation studies: we evaluate how reducing the size of the synthetic pre-training dataset affects tensorial prediction performance.
We consider four series of \texttt{TensorMACE}-$P$-$F$(L) models characterized by the size of the pre-training set $P$: 10, 100, 1000, and 5000 synthetically labeled structures. 
The performance of these models is indicated by the shades of blue in Fig.~\ref{fig:sio2}. For any given number $F$ of QM fine-tuning structures, we find that the accuracy of the model increases with the size of the pre-training set, albeit with diminishing returns beyond 1000 structures. This behavior suggests that the dominant correlations between local geometry and response are already captured with a few hundred synthetic structures, so additional pre-training data yield progressively smaller gains.

For a more complete assessment of the data efficiency of synthetic pre-training, we also train a \texttt{TensorMACE} model directly, using 800 a-\ce{SiO2} structures, corresponding to the training set used in previous work~\cite{benmahmoud_graphneuralnetwork_2025}. We find that the performance of a direct model \texttt{TensorMACE}-0-800(L), which uses the entire training set of a previous work~\cite{benmahmoud_graphneuralnetwork_2025},  is comparable to that of \texttt{TensorMACE}-5000-250(L), which corresponds to a threefold reduction in the number of QM labels required. In particular, this direct model achieves \%RMSE values of 6.7\% (silicon) and 8.0\% (oxygen) for $\sigma^{(0)}$, 9.2\% (silicon) and 29.7\% (oxygen) for $\sigma^{(1)}$, 8.2\% (silicon) and 6.7\% (oxygen) for $\sigma^{(2)}$.

\subsection{Quality of pre-training labels}

To assess the sensitivity to the quality of the pre-training labels, we consider a scenario in which imperfect synthetic supervision is available, as may arise when synthetic labels are generated from small datasets, lower-accuracy QM methods, or transferable (ML) models.
We train a second NequIP-based model using only 50 a-\ce{SiO2} configurations. This model achieves RMSE values of 2.0~ppm for silicon MS isotropy and 3.8~ppm for the oxygen MS isotropy, indicating that the resulting synthetic labels are significantly less accurate than those used in Sec.~\ref{sec:pre-train-sio2}. Using this model, we generate synthetic labels for the 5,000 structures and pre-train a \texttt{TensorMACE}(L) using the same protocol as previously. We then fine-tune the model on increasing numbers of a-\ce{SiO2} structures ranging from 10 to 250. 

Figure~\ref{fig:noisy} compares models pre-trained with higher- and lower-quality synthetic labels, averaged over 4 independent training runs. Models pre-trained with higher-quality data consistently achieve lower errors in the low-data regime ($<50$ structures), while the two pre-training strategies start to converge as the amount of QM training data increases to 250 structures. Importantly, pre-training with lower-quality labels remains competitive with direct training: the best directly trained model considered, trained on 250 QM-labeled structures, achieves a performance comparable to that of the lower-quality pre-trained model fine-tuned on 50 structures. This corresponds to a five-fold reduction in the number of QM-labels required for training. 
The result suggests that the benefit of synthetic pre-training is not simply due to the number of QM-labeled structures used to train the synthetic label generator.

\subsection{Model capacity}
\begin{figure*}
    \centering
    \includegraphics[width=1.\linewidth]{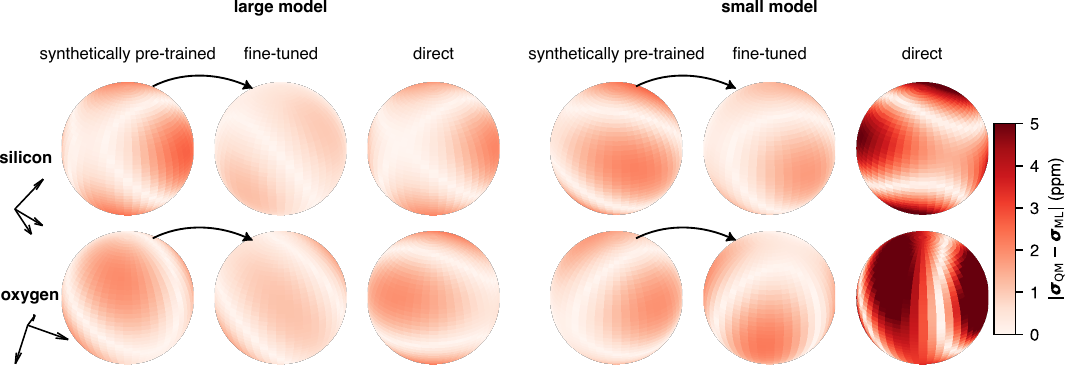}
    \caption{Orientation-resolved deviations of the predicted $\sigma^{(2)}$ components compared to the QM ground truth shown as spherical maps. Rows correspond to element: silicon, oxygen. Columns compare the synthetically pre-trained, fine-tuned, and directly trained large and small models. Maps correspond to the atom with the median test set error of the large fine-tuned model. The basis axes correspond to the principal axes of the QM tensor.}
    \label{fig:ellipsoids}
\end{figure*}
Next, we use the same pre-training and fine-tuning protocol to examine the role of model capacity. We train a series of reduced-size \texttt{TensorMACE}-$P$-$F$(S) models, where $P=$ 10, 100, 1000, or 5000, and compare them with the corresponding directly-trained \texttt{TensorMACE}-0-$F$(S) models. We also compare to the best-performing large model \texttt{TensorMACE}-5000-250(L). The corresponding architectural hyperparameters are reported in Sec.~\ref{sec:fitting}. In particular, the small and large models differ only in the maximum rank of their equivariant features.

 Figure~\ref{fig:sio2_small} shows that, for any pre-training set size and number of QM-labeled structures, the \texttt{TensorMACE}-$P$-$F$(S) models exhibit larger errors than their large counterparts, particularly for the anisotropic quantities. Increasing the number of pre-training labels improves the performance of the models, yet, the best small model is less accurate than a large \texttt{TensorMACE}-5000-25(L). This indicates that synthetic pre-training partially compensates for the reduced model capacity by guiding the optimization toward physically relevant features. The comparison with direct training highlights this effect even more clearly. For example, \texttt{TensorMACE}-0-250(S) performs substantially worse than \texttt{TensorMACE}-1000-250(S), even though both are trained or fine-tuned on the same 250 QM-labeled structures. In particular, the relative errors of the asymmetry parameter $\sigma_\eta$ for silicon exceed 100\% for each one of the directly trained \texttt{TensorMACE}-0-F(S). Hence, our results suggest that synthetic pre-training improves the accuracy of the small architecture and the stability of the training.
 
 Increasing the number of pre-training structures to 5000 improves the performance of the small models, but only up to the expressivity limit of the architecture. For $P<$5000, the \texttt{TensorMACE}-$P$-$F$(S) series benefit from additional QM fine-tuning labels. However, the \texttt{TensorMACE}-5000-$F$(S) series quickly reaches a plateau in the low-data regime as indicated in Fig.~\ref{fig:sio2_small}. This indicates that the small architecture becomes capacity-limited, and additional QM fine-tuning configurations provide only limited further improvement.

\subsection{Tensor orientation accuracy}

Figures~\ref{fig:sio2} and \ref{fig:sio2_small} show that synthetic pre-training improves the fidelity of predictions for the rank-2 components, $\sigma^{(2)}$, which are responsible for the symmetric traceless part of the MS tensor, and that this improvement translates into better predictions of derived anisotropy parameters.  
The large \texttt{TensorMACE}-5000-250(L) model, and its directly trained analogue, \texttt{TensorMACE}-0-250(L), achieve comparable errors on $\sigma^{(2)}$: for silicon, 7.6\% vs 12.1\% \%RMSE, and for oxygen, 7.0\% vs 11.2\% \%RMSE. The corresponding small models show a larger gap between fine-tuned and directly-trained models: for silicon, 15.3\% vs 35.3\% \%RMSE, and for oxygen, 11.8\% vs 26.7\% \%RMSE.
These results indicate that synthetic pre-training narrows the performance gap between small and large architectures for the spherical tensor components.

The effect becomes more pronounced when considering the Haeberlen asymmetry parameter $\sigma_\eta$. For silicon, the errors increase to 29.3\% (large, fine-tuned) and 70.0\% (small, fine-tuned), whereas the directly trained counterparts yield 45.2\% (large) and 110.2\% (small).
For oxygen, the asymmetry is similarly more stable if pre-training is used: fine-tuned large and small models yield 1.7\% and 34.7\% \%RMSE, respectively, versus 35.8\% and 67.6\% for the directly trained models. These comparisons show that the anisotropy measures are more sensitive to small $\sigma^{(2)}$ component errors, and that synthetic pre-training reduces these discrepancies, especially for larger models.

To assess the fidelity of the predicted symmetric tensor resulting from the  $\sigma^{(2)}$ components beyond scalar metrics, we examine the orientation dependence of $\sigma^{(2)}$. Although the ML models predict the spherical components, we first reconstruct the traceless symmetric tensor in its Cartesian representation to evaluate the anisotropic contribution to $\boldsymbol{\sigma}$ along a direction $\hat{\mathbf{n}}$:
$$
\sigma^{(2)}(n) = \hat{\mathbf{n}}^\top \sigma^{(2)} \hat{\mathbf{n}}.
$$
Figure~\ref{fig:ellipsoids} presents spherical maps of the absolute deviation between the ML predictions and the QM ground truths, evaluated over the unit sphere. For both silicon and oxygen, the maps compare synthetically pre-trained, fine-tuned, and directly trained models, and are shown for the large and small ML architectures. The examples selected correspond to the atom associated with the median prediction error of the large synthetically pre-trained (on 5000 structures) and then fine-tuned (on 250 structures) model for silicon and oxygen. 

For the large architecture, the deviations remain relatively small and broadly distributed across the entire sphere for both silicon and oxygen and across all training protocols. The fine-tuned model yields the lowest directional discrepancies, while the directly-trained model exhibits slightly larger orientation-dependent errors. 
The synthetically pre-trained model, prior to fine-tuning (\texttt{TensorMACE}-5000-0(L)), already captures the main directional features of the MS tensor, with only minor residual deviations. 
We also notice that errors are the lowest along the principal axes of the MS tensor, indicating that the prediction of the eigenvalues of $\boldsymbol{\sigma}$ is largely preserved, while the residual errors arise from the small misalignments of the principal axes between ML predictions and QM ground truth and distortions of the anisotropic shape.
For the small architecture, the differences between training strategies are more pronounced. The directly-trained model shows strong localized regions of large deviations for both silicon and oxygen. Synthetic pre-training captures the orientation-dependent information relatively well, and fine-tuning further improves the uniformity of the deviations, bringing the small model closer to the behavior of the large architecture.

While this analysis does not constitute a statistical characterization of the orientation-resolved errors across the test set, it provides a representative illustration of how differences in the learned anisotropic components manifest as directional discrepancies of the MS tensor. The trends observed in the maps of Fig.~\ref{fig:ellipsoids} are consistent with the scalar error metrics reported for the full test set as discussed in Sec.~\ref{sec:pre-train-sio2}, and therefore offer a qualitative visualization reflecting the systematic trends captured by the dataset-level analysis. An accurate prediction of the orientation of the MS tensor is often essential for reproducing the NMR observables, particularly when multiple interactions are present and their mutual orientation influences the spectrum. Even for a single interaction, the resonance frequency depends on the orientation of the tensor with respect to the external magnetic field, and the resulting powder line shape arises from the distribution of these orientation-dependent contributions~\cite{duer_introduction_2004}. 

\begin{figure*}
    \centering
    \includegraphics[width=1.\linewidth]{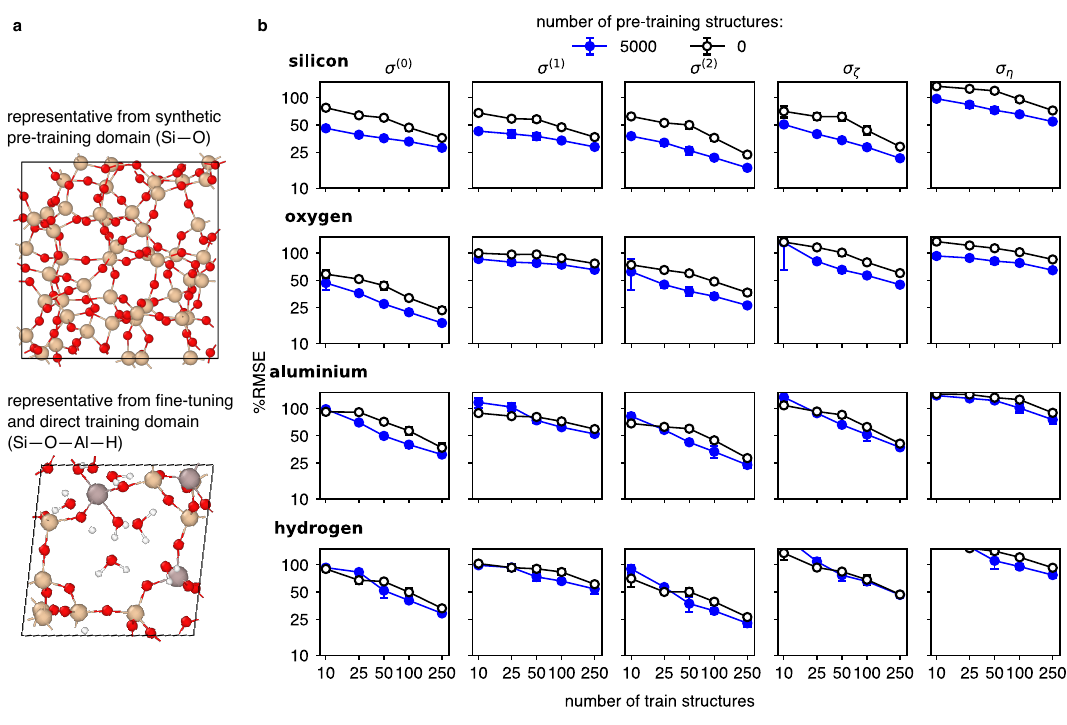}
    \caption{Transferability of synthetically pre-trained models from a-\ce{SiO2} to Si--O--Al--H zeolite systems. 
    (a) Representations of typical configurations in the fine-tuning datasets. Upper: a-\ce{SiO2}. Lower: zeolite with a water molecule in the cavity. Renders performed with OVITO~\cite{stukowski_visualization_2010}.
    (b) Prediction errors for isotropic and anisotropic components of the MS tensor for silicon, oxygen, aluminium, and hydrogen as a function of the QM NMR training structures. The synthetic pre-training is performed exclusively on a-\ce{SiO2}.
    Blue symbols refer to the best-performing synthetically-pre-trained-and-fine-tuned large \texttt{TensorMACE} model. Empty circles correspond to models trained directly.
    Error bars indicate the standard deviation of the prediction errors across four models trained with different random seeds. When not visible, the standard deviation is smaller than the marker size.}
    \label{fig:sioalh}
\end{figure*}

\subsection{From a-SiO$_2$ to Si--O--Al--H systems}\label{sec:pre-train-sioalh}

\begin{figure*}
    \centering
    \includegraphics[width=1.\linewidth]{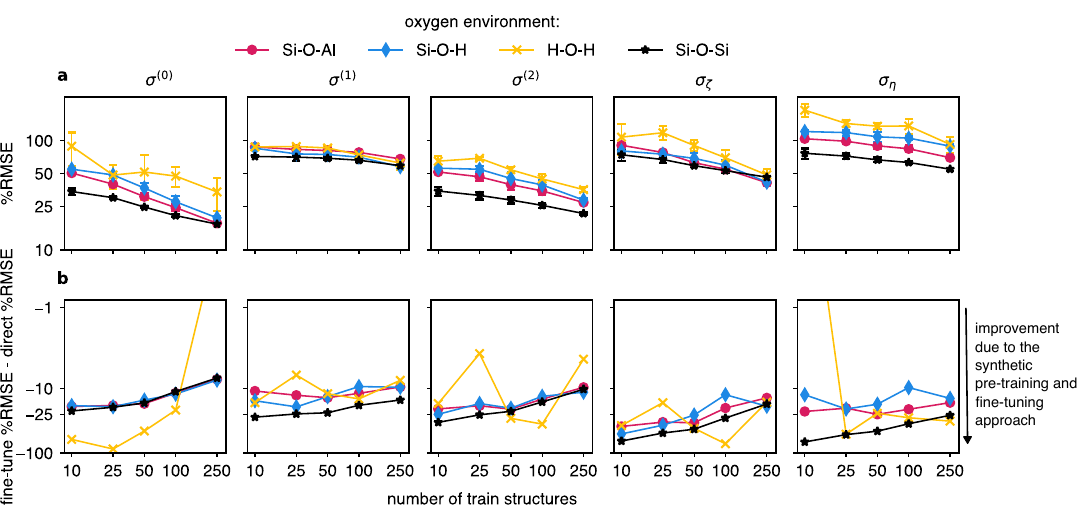}
    \caption{Transferability of synthetically pre-trained models from a-\ce{SiO2} to Si--O--Al--H zeolite systems. 
    We show the prediction errors for isotropic and anisotropic components of the MS tensor for oxygen as a function of the QM NMR training structures used for fine-tuning (first row) and differences in prediction errors between the fine-tuned and direct models (second row). The synthetic pre-training is performed exclusively on a-\ce{SiO2}. 
    Red circles refer to the Si--O--Al environments including Q$^2$(0Al), Q$^3$(0Al) and Q$^3$(1Al)~\cite{greiser_29si27al_2016}, blue diamonds to Si--O--H environments including bridging hydroxyl and their corresponding deprotonated counterparts, yellow crosses to oxygen atoms sharing a proton, and black stars to Si--O--Si environments.
    Error bars indicate the standard deviation of the prediction errors across four models trained with different random seeds. When not visible, the standard deviation is smaller than the marker size.}
    \label{fig:oenvs_sioalh}
\end{figure*}
Finally, we examine the transferability of synthetically pre-trained tensor models to a chemically more diverse system containing elements that are absent during the synthetic pre-training phase. Representative structures from both datasets are visualized in Fig.~\ref{fig:sioalh}(a).

We begin by assessing the compatibility between MS tensors calculated with the PBE functional (used for synthetic pre-training), and PBEsol (the reference for the zeolites), in a small dataset of 25 zeolitic structures from the training set. We find that the difference between the exchange--correlation functionals is dominated by a nearly constant shift in the isotropic component $\sigma^{(0)}$, with minimal impact on the anisotropy of the tensors. 
In practice, the near-constant shift in $\sigma^{(0)}$ does not affect the fine-tuning of the synthetically pre-trained \texttt{TensorMACE} models, as discussed in Sec.~\ref{sec:fitting}. Because the model includes a learnable per-element offset and the data are mean-centered prior to training, these shifts in the spherical components are absorbed into the offset term.
Quantitatively, the discrepancy in $\sigma^{(0)}$ between PBE and PBEsol for hydrogen is 0.3~ppm, for oxygen is 5.4~ppm, for aluminium is 3.8~ppm, and for silicon is 4.1~ppm. The shifts in  $\sigma^{(1)}$ are less than 0.2~ppm for all nuclei and in $\sigma^{(2)}$ are less than 0.3~ppm for hydrogen, aluminium, and silicon and 0.8~ppm for oxygen. 

Figure~\ref{fig:sioalh}(b) reports prediction errors for the different isotropic and anisotropic parameters of the MS tensor obtained when a large \texttt{TensorMACE} model is pre-trained on 5000 configurations of a-\ce{SiO2} and then fine-tuned on a dataset of zeolitic structures containing silicon, oxygen, aluminium, and hydrogen. Despite the absence of Al and H in the synthetic pre-training data, the pre-trained model consistently outperforms the directly trained model in the low-data regime ($< 100$ QM-labeled training structures). 
The improvement is most pronounced for silicon and oxygen, which are present during pre-training. Similar to the a-\ce{SiO2} case, the largest gains are observed for anisotropic tensor components. This observation suggests that the synthetic pre-training encodes local, partially transferable, geometric features that remain relevant across chemical compositions, while finer element-specific effects are learned during fine-tuning. For aluminium and hydrogen, the benefit is smaller, as both models reach comparable performance for all QM-labeled structures in their training and fine-tuning sets. To place the improvements linked to synthetic pre-training in perspective, we report, in Table~\ref{tab:sioalh}, the average RMSE values for silicon, oxygen, aluminium and hydrogen in this dataset.

\begin{table}
\caption{\label{tab:sioalh}Average RMSE values (in ppm) for the Si--O--Al--H dataset at 250 QM-labeled training structures. We compare the synthetically pre-trained and fine-tuned large \texttt{TensorMACE} model with the corresponding directly trained model.}
\begin{ruledtabular}
\begin{tabular}{lcccc}
Element & \multicolumn{2}{c}{$\sigma^{(0)}$} & \multicolumn{2}{c}{$\sigma^{(2)}$} \\
\cline{2-3}\cline{4-5}
 & Fine-tuned & Direct & Fine-tuned & Direct \\
\hline
Si & 5.1 & 6.6 & 5.9 & 8.3 \\
O  & 8.9 & 12.3 & 8.3 & 11.6 \\
Al & 4.8 & 5.7 & 9.0 & 10.7 \\
H  & 1.3 & 1.5 & 1.9 & 2.2 \\
\end{tabular}
\end{ruledtabular}
\end{table}

Despite silicon and oxygen being present during pre-training, the Si--O--Al--H system contains a much more diverse chemical and configurational space for these atoms. 
Looking at the oxygen environments, the chemical diversity increases from Si--O--Si motifs to Si--O--Al, Si--O--H, and protonated oxygen species. Figure~\ref{fig:oenvs_sioalh} shows the prediction errors of the fine-tuned model for the different isotropic and anisotropic quantities of the MS tensor across these newly-encountered oxygen environments. For reference, we also include the performance on the Si--O--Si environments of the zeolitic dataset. We also compare to a directly trained model. 

As expected, the environments that closely resemble those seen during pre-training, such as Si--O--Si and Si--O--Al with local geometries similar to Si--O--Si environments in the pre-training set, show the largest gains from synthetic pre-training on a-\ce{SiO2}. However, more distant environments, such as H--O--H, benefit less strongly but still achieve typical performance gains of 10\%--25\% when evaluating the \%RMSE metric relative to direct training.
These results indicate that the model learns transferable geometry–tensor correlations that generalize beyond the specific chemistry of a-\ce{SiO2}. This behavior is consistent with the spatial character of the shielding response: for second-period elements such as oxygen, the dominant contributions arise from relatively localized electronic currents, while for hydrogen, the shielding is more strongly influenced by long-range currents~\cite{zilka_visualising_2017}. As a result, the (semi)-local geometric descriptors of GNNs, due to their message-passing formulation, are expected to transfer more readily across oxygen environments, while accurately predicting hydrogen shielding requires capturing longer-range interactions.

\section{Conclusions and outlook}

We have shown that synthetic pre-training can improve the accuracy of tensorial ML models targeting NMR parameters in regimes where high-quality training data from direct quantum-mechanical computations are scarce. 
Using a-\ce{SiO2} as a prototypical system, we showed that synthetic pre-training in an ``in-domain'' setting improves learning across both isotropic and anisotropic components of the MS tensor. Exposure to synthetic labels biases the model toward learning geometry--tensor correlations prior to first-principles supervision. This leads to substantial gains in data efficiency, ranging between threefold and tenfold in our experiments depending on the data scarcity regime of the direct approach, while leaving the asymptotic accuracy in the data-rich regime largely unchanged.

Extending this approach across chemical domains, from a-\ce{SiO2} to a more complex Si--O--Al--H zeolitic system, reveals that this transfer is inherently partial. Pre-training on a-\ce{SiO2} provides the strongest benefits when predicting tensorial properties for chemical species that have been encountered during pre-training, but also yields measurable improvements for previously ``unseen'' environments in the low-data regime, as long as the local geometry is close to that of environments in the pre-training set. This behavior indicates that synthetic pre-training predominantly encodes geometry-driven correlations that are partially transferable across chemical species, while element-specific (electronic) effects are refined through first-principles supervision.

The dependence of the performance on both pre-training data volume and architectural capacity provides further insight into the role of synthetic data. Once the dominant geometric features within a given chemical domain have been learned, additional synthetic data yield diminishing returns, suggesting that pre-training serves to establish an approximate representation of the tensorial target. At the same time, pre-training mitigates, but does not eliminate, the limitations imposed by reduced model expressivity, highlighting its role as a guide for optimization rather than a substitute for architectural capacity.

Looking further ahead, our work motivates the development of a more general, automated workflow for the efficient training of tensorial ML models. As mentioned in the Introduction and in Ref.~\citenum{bornes_accurate_2026}, an important difference between MLIP and NMR models in this context is in the availability of data. Energy and force labels for MLIPs can be systematically computed with first-principles methods, such as DFT, for almost any configuration -- in contrast, NMR parameters are not always accessible, and even when they are, the computation is often significantly more demanding. 
Hence, pre-trained (``foundational'') MLIPs~\cite{deng_chgnet_2023,neumann_orb_2024,batatia_foundation_2025,mazitov_petmad_2025} can be used to sample large regions of chemically relevant space at low cost, while approximate or transferable NMR models provide inexpensive yet physically informed synthetic tensorial labels for pre-training. Such models do exist for molecular crystals~\cite{kellner_deep_2025} and proteins~\cite{han_shiftx2_2011};
however, to our knowledge, comparably comprehensive ML-based approaches have not yet been developed for inorganic systems, representing an important direction for future work. High-accuracy, first-principles NMR computations can then be reserved for targeted fine-tuning in the specific region of interest, providing chemical specificity and quantitative accuracy where required~\cite{kuryla_how_2025}. By shifting the computational effort away from exhaustive first-principles data generation and toward synthetic pre-training followed by focused refinement, such workflows could offer scalable routes to accurate broadly transferable ML-driven NMR models for inorganic materials.

\section*{Data and code availability}
Data supporting this work, including train and test sets, synthetic labels, \texttt{graph-pes} input files, and ML models weights can be found at \href{https://github.com/cbenmahm/synthetic-nmr}{https://github.com/cbenmahm/synthetic-nmr}. 
The \texttt{graph-pes} software is available at \href{https://github.com/vldgroup/graph-pes}{https://github.com/vldgroup/graph-pes}.

\begin{acknowledgments}
We thank J.~L.~A.~Gardner for help with implementing the \texttt{TensorMACE} models in \texttt{graph-pes}. 
CB, LG and CJH acknowledge the support of Czech Science Foundation (26-23277S) and Charles University Centre of Advanced Materials (CUCAM, OP VVV Excellent Research Teams, project number CZ.02.1.01/0.0/0.0/15\_003/0000417). The computations at Charles University were supported by the Ministry of Education, Youth and Sports of the Czech Republic through the e-INFRA CZ (ID:90254). CB acknowledges the funding from the European Union's Horizon Europe research and innovation program under the ERA-PF grant agreement no. 101180584.
This work was supported by UK Research and Innovation [grant number EP/X016188/1].
\end{acknowledgments}
\bibliographystyle{aipnum4-2}
%aipnum4-2.bst 2019-01-14 (MD) hand-edited version of apsrev4-1.bst
%Control: key (0)
%Control: author (8) initials jnrlst
%Control: editor formatted (1) identically to author
%Control: production of article title (-1) disabled
%Control: page (0) single
%Control: year (1) truncated
%Control: production of eprint (0) enabled
%

\end{document}